\documentclass[a4paper,fleqn,usenatbib]{mnras}

\usepackage{newtxtext,newtxmath}
\usepackage{tablefootnote}

\usepackage[T1]{fontenc}
\usepackage{ae,aecompl}
\usepackage{ulem}

\usepackage{graphicx}	
\usepackage{amsmath}	
\usepackage{amssymb}	

\title[A survey of hot Jupiters in secondary eclipse]{A survey of eight hot Jupiters in secondary eclipse using WIRCam at CFHT}

\author[E. Martioli et al.]{
Eder Martioli$^{1}$\thanks{E-mail: emartioli@lna.br},
Knicole D. Col\'on$^{2,3,17}$,
Daniel Angerhausen$^{4,16}$,
Keivan G. Stassun$^{5,6}$, \newauthor
Joseph E. Rodriguez$^{5,7}$,
George Zhou$^{7}$,
B. Scott Gaudi$^{8}$,
Joshua Pepper$^{9}$, \newauthor
Thomas G. Beatty$^{10,11}$,
Ramarao Tata$^{12}$,
David J. James$^{13}$, 
Jason D. Eastman$^{7}$,  \newauthor
Paul Anthony Wilson$^{14}$,
Daniel Bayliss$^{15}$,
Daniel J. Stevens$^{5}$ \\
\\
$^{1}$Laborat\'{o}rio Nacional de Astrof\'{i}sica (LNA/MCTI),  Rua Estados Unidos 154, Itajub\'{a}, MG, Brazil\\
$^{2}$NASA Ames Research Center, M/S 244-30, Moffett Field, CA 94035, USA\\
$^{3}$Bay Area Environmental Research Institute, 625 2nd St. Ste 209 Petaluma, CA 94952, USA\\
$^{4}$USRA NASA Postdoctoral Program Fellow, NASA Goddard Space Flight Center, 8800 Greenbelt Road, Greenbelt, MD 20771,
USA\\
$^{5}$Department of Physics and Astronomy, Vanderbilt University, 6301 Stevenson Center, Nashville, TN 37235, USA\\
$^{6}$Department of Physics, Fisk University, Nashville, TN 37208, USA \\
$^{7}$Harvard-Smithsonian Center for Astrophysics, 60 Garden St, Cambridge, MA 02138, USA\\
$^{8}$Department of Astronomy, The Ohio State University, 140 West 18th Avenue, Columbus, OH 43210, USA\\
$^{9}$Department of Physics, Lehigh University, 16 Memorial Drive East, Bethlehem, PA 18015, USA\\
$^{10}$Department of Astronomy \& Astrophysics, The Pennsylvania State University, 525 Davey Lab, University Park, PA 16802, USA\\
$^{11}$Center for Exoplanets and Habitable Worlds, The Pennsylvania State University, 525 Davey Lab, University Park, PA 16802, USA\\
$^{12}$Ohio University, Athens, OH 45701, USA \\
$^{13}$Department of Astronomy, University of Washington, Box 351580, Seattle, WA 98195, USA \\
$^{14}$CNRS, UMR 7095, Institut d'Astrophysique de Paris, $98^{\mathrm{bis}}$ Boulevard Arago, F-75014 Paris, France\\
$^{15}$Observatoire Astronomique de l'Universit\'e de Ge\`eve, Chemin des Maillettes 51, 1290 Sauverny, Switzerland\\
$^{16}$Center for Space and Habitability, University of Bern, Sidlerstrasse 5, 3012 Bern, Switzerland  \\
$^{17}$NASA Goddard Space Flight Center, Exoplanets and Stellar Astrophysics Laboratory (Code 667), Greenbelt, MD 20771, USA \\
}
\date{Accepted 2017 November 18. Received 2017 November 17; in original form 2016 July 28}

\pubyear{2017}

\begin{document}
\label{firstpage}
\pagerange{\pageref{firstpage}--\pageref{lastpage}}
\maketitle

\begin{abstract}

We present near infrared high-precision photometry for eight transiting hot Jupiters observed during their predicted secondary eclipses. Our observations were carried out using the staring mode of the WIRCam instrument on the Canada-France-Hawaii Telescope (CFHT). We present the observing strategies and data reduction methods which delivered time series photometry with statistical photometric precision as low as 0.11\%. We performed a Bayesian analysis to model the eclipse parameters and systematics simultaneously. The measured planet-to-star flux ratios allowed us to constrain the thermal emission from the day side of these hot Jupiters, as we derived the planet brightness temperatures. Our results combined with previously observed eclipses reveal an excess in the brightness temperatures relative to the blackbody prediction for the equilibrium temperatures of the planets for a wide range of heat redistribution factors.  We find a trend that this excess appears to be larger for planets with lower equilibrium temperatures.  This may imply some additional sources of radiation, such as reflected light from the host star and/or thermal emission from residual internal heat from the formation of the planet.  

\end{abstract}



\section{Introduction}

A secondary eclipse of an exoplanet occurs when the planet passes behind the host star as it moves along its orbit.  The star blocks the light emerging from the planet's atmosphere, allowing direct measurements of the planet-to-star flux ratio. Infrared observations of these eclipses provide constraints on the thermal emission of the exoplanet, which allows a more detailed characterization of its atmosphere.  A number of eclipses have been measured in the mid-infrared bands using the Spitzer/IRAC instrument \cite[e.g.,][]{kammer2015} and also in the near infrared bands from ground-based observations \cite[e.g.,][]{croll2010a, croll2015, zhou2014, zhou2015}. These observations allowed important studies in the characterization and understanding of hot Jupiter atmospheres \cite[e.g.,][]{harrington2007,knutson2007,charbonneau2008,knutson2009,stevenson2010,knutson2011}.

Gathering a large sample of hot Jupiters with measured secondary eclipses is important in order to perform a comparative analysis of these objects over a range of exoplanet parameters and in different stellar environments. \citet{zhou2015} presented near infrared secondary eclipse observations for seven hot Jupiters, and in addition they presented a compilation of all other detections of secondary eclipses from both ground-based observations in the near infrared and from the four Spitzer IRAC bandpasses.  As this sample grows, it becomes possible to place statistically significant constraints on the atmospheric models of hot Jupiters, as done in several consistent comparative studies of secondary eclipses using optical \textit{Kepler} time series \citep[e.g.,][]{2013ApJ...772...51E,henganddemory2013,2015PASP..127.1113A,schwartz2015}. In this paper, we present new results that add to the previous literature on infrared secondary eclipse measurements.

The Wide-field InfraRed Camera \cite[WIRCam,][]{puget2004} at the Canada-France-Hawaii Telescope (CFHT) is one of the primary ground-based instruments capable of detecting thermal emission of exoplanets through the observation of secondary eclipses \cite[e.g.,][]{croll2010a, croll2010b}. CFHT uses an observing mode called ``staring mode", where the observations are performed without dithering, keeping the field at a constant position in the detector. The position is kept constant at the level of a few pixels (with drifts no larger than 2 pixels), which is an important factor in achieving high precision photometry necessary to detect small eclipse signals.  

\citet{croll2015} presented detections of secondary eclipses for three hot Jupiters and one brown dwarf using WIRCam on the CFHT obtained using the so called ``staring mode'' \citep{croll2010a}. In their paper, they stressed the importance of obtaining robust and repeatable measurements of eclipses from ground-based observations. This is challenging given the nature of timed observations of eclipses, since the eclipses do not necessarily occur always during the most suitable observing conditions. They also presented revised methods for the reduction of CFHT/WIRCam staring mode data, where they obtained photometric precisions (RMS of the residuals per exposure) between 0.15\% and 0.40\%. However, \citet{devost2010} estimated the signal-to-noise ratio obtained using the staring mode in ideal conditions and compared their results with measurements of the eclipse of the hot Jupiter TrES-2b in K-band, where they concluded that despite the high precision attained in their CFHT/WIRCam experiment, their measurements were not photon noise limited.  
We performed observations with the aim of measuring near infrared eclipses for a number of hot Jupiters also using WIRCam in staring mode, where we adopted a similar technique as in \citet{croll2015}. However, in our work we present a more detailed investigation of the different factors that affect the photometric precision. In \autoref{sec:observations} we present the observations and the experimental design adopted in our experiment, where we discuss the aspects concerning the telescope defocus and exposure time.  We also present a study on the detector characteristics, which are further considered in the reduction of our data, such as in the identification and removal of bad pixels, and the flat-field and non-linearity corrections. In \autoref{sec:datareduction} we present a description of data reduction procedures used in our pipeline, where we introduce some new ideas to mitigate, identify, and remove the systematics.  Also in \autoref{sec:datareduction} we present a Bayesian analysis of our light curves to obtain the system's parameters. Finally in \autoref{sec:discussion} we discuss our results and some possible implications when combining them with previously published measurements of eclipses of hot Jupiters.

\section{Observations and Experimental Design}
\label{sec:observations}

We were awarded 25 hours in 2014\,A and 17 hours in 2014\,B of telescope time at CFHT in queue mode (programs 14AB99, 14AB01, 14BB98 and 14BB02). With the time awarded we were able to observe a sample of eight transiting hot Jupiters in eclipse with a wide range of host star and planetary parameters. \autoref{table:obslog} shows a summary of all eclipse events observed as part of these programs. All events were observed in photometric conditions. The observations typically started about an hour before the predicted start of eclipse and ended about an hour after the end of eclipse. We prioritized observations in the K-band (at $\lambda\sim 2.2\,\mu$m), where the near infrared contrast between the star and a typical hot Jupiter is optimized. We used either broad band or narrow band filters (see \autoref{table:obslog}), depending on the target magnitude, to avoid saturation of the detector.  Also included in this table is WASP-12, which was previously observed by \citet{croll2015} and which we used as a test case in order to compare our derived photometry with the previous state of the art from CFHT \citep{croll2015}.

\begin{table*}
\caption{Log of observations. RA and DEC data are from the 2MASS catalog \citep{skrutskie2006}.}
\label{table:obslog}
\begin{tabular}{cccccccc}
\hline
Object ID & RA(2000) & DEC(2000) & Filter $^{1}$ & UT Date & Time span & Exptime & Defocus\\
 & hh:mm:ss.ss & $\pm$dd:mm:ss.ss &  &  & (h) & (s) & (mm) \\
\hline
\hline
WASP-12 & 06:30:32.79 & +29:40:20.25 &  Ks & 2009-12-28 & 6.39 & 5.0  & 2.0  \\
KELT-4A & 10:28:15.01 & +25:34:23.60 &  Ks  & 2014-03-19 & 4.98 & 4.0 & 1.8  \\
WASP-14 & 14:33:06.35 & +21:53:40.97 &  CH4Off & 2014-05-18 & 5.0 & 5.0 & 2.0  \\
TrES-4 & 17:53:13.04 & +37:12:42.66 &  Ks  & 2014-07-13 & 2.71 & 5.0 & 1.5  \\
Kepler-5 & 19:57:37.68 & +44:02:06.17 &  Ks  & 2014-08-03 & 8.84 & 12.0 & 1.5  \\
KELT-2A & 06:10:39.35 & +30:57:25.86 & Kcont & 2014-12-09 & 7.71 & 6.0 & 1.5   \\
KELT-7 & 05:13:10.93 & +33:19:05.41 &  Kcont & 2014-12-10 & 5.93 & 8.0 & 1.5    \\
WASP-31 & 11:17:45.37 & -19:03:17.18 &  Ks  & 2015-01-31 & 5.15 & 6.0 &  1.5 \\
HAT-P-33 & 07:32:44.21 & +33:50:06.19 &  Ks  & 2015-02-01 & 4.32 & 6.0 & 1.5 \\
\hline
\end{tabular}
\end{table*}

\subsection{Telescope defocus and exposure time}
\label{sec:defocusandexptime}

In order to test the best instrumental set up for our observations, we performed several preliminary observations (pre-imaging) using different amounts of defocus and exposure times for several filters. These observations allowed us to evaluate the effect of defocus on the Point Spread Function (PSF) and to select the best instrument configuration. Telescope defocus is commonly used in staring mode observations of WIRCam to improve the photometric precision \cite[e.g.,][]{croll2010a, croll2010b, croll2015}. The amount of defocus used in previous work typically ranged between 1.0 and 2.0\,mm.  The defocusing is intended to avoid saturation and also to reduce the effects of bad pixels and flat-fielding errors. In defocused images the star PSF is broadened, spreading the light over a larger area of the detector, reducing the concentration of light in a few central pixels, which also decreases the amount of light affected by a single defective pixel.  However, the more pixels used the more readout and background noise is added to the measurement, resulting in a lower Signal-to-Noise Ratio (SNR). The lower SNR can be compensated by increasing the exposure time, which also decreases the cadence of the observations.

\begin{figure}
\includegraphics[width=\columnwidth]{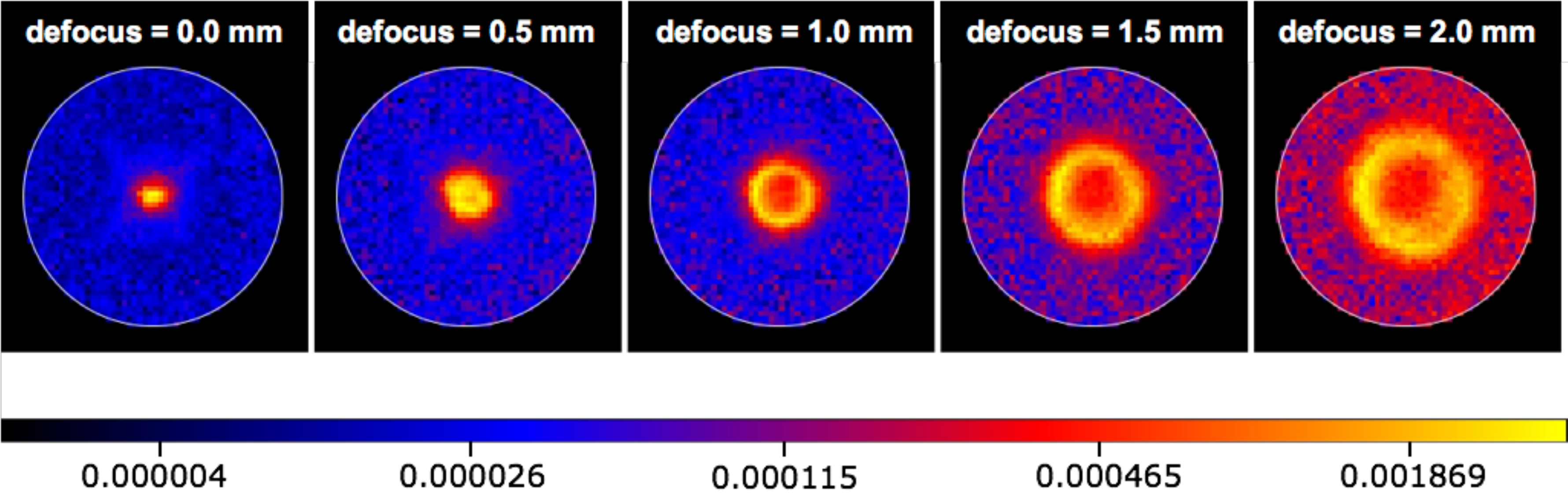}
\caption{WIRCam PSF for five different levels of telescope defocus. From left to right the panels show defocus of 0.0, 0.5, 1.0, 1.5, and 2.0 milimiters, respectively. The illumination fraction is presented in colour code as indicated by the colour bar, which ranges from 0.0004\% (dark blue) to 0.2\% (bright yellow).} \label{fig:DefocusPSFWithCirclesInLine}
\end{figure}

\begin{figure}
\centering
\includegraphics[width=\columnwidth]{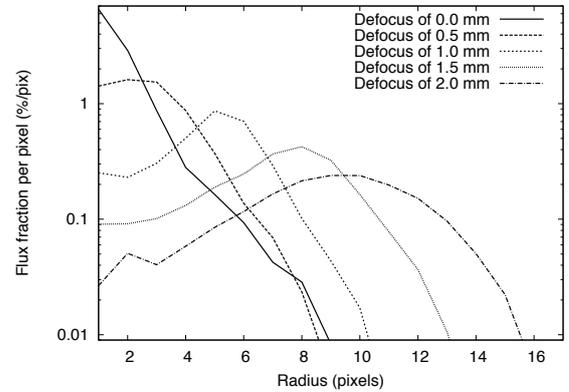} 
\caption{Radial profile measured on the WIRCam PSF for 5 different levels of defocus as indicated in the legend.} \label{fig:radProfileFFperPixel}
\end{figure}

 Here we present results obtained from several images taken on two different nights from the fields of KELT-2A and KELT-7, and for several combinations switching between narrow band filters (KCont, CH4Off, and LowOH2)\footnote{\label{foot:wircamch4offfilter} For more information on WIRCam filters see: \url{http://www.cfht.hawaii.edu/Instruments/Filters/wircam.html}}, telescope defocus (from 0.0 to 2.0 mm), and exposure times (from 3.0 to 12.0 seconds).   \autoref{fig:DefocusPSFWithCirclesInLine} shows an example of the WIRCam PSF measured from several stars in the field of KELT-2A, where we defocused the telescope by an amount varying between 0.0 and 2.0\,mm in steps of 0.5\,mm.  Notice that as one moves out of focus the PSF becomes a ``doughnut-like'' shape, which is a direct function of the pupil image.  \autoref{fig:radProfileFFperPixel} shows radial profiles for the flux fraction divided by the number of pixels within each annulus of 1 pixel width. This shows the concentration of flux per pixel at different apertures.

\begin{figure}
\includegraphics[width=\columnwidth]{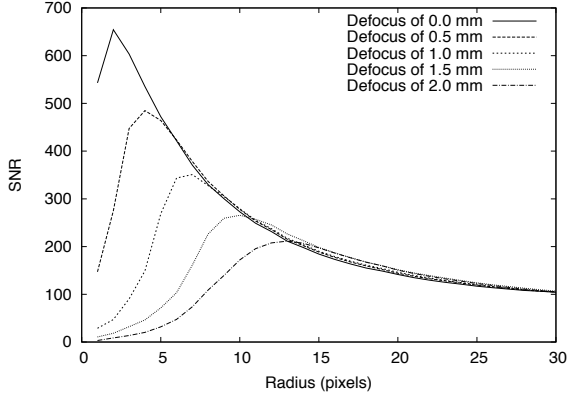}
\caption{Signal-to-noise ratio versus aperture radius for simulated data using the measured WIRCam PSF for 5 different amounts of defocus, ranging from 0.0 to 2.0\,mm, and injecting a star signal of $10^6$\,photons and sky background flux of 28,000\,photons/pixel.}  \label{fig:radProfileSNR}
\end{figure}

\begin{figure}
\includegraphics[width=\columnwidth]{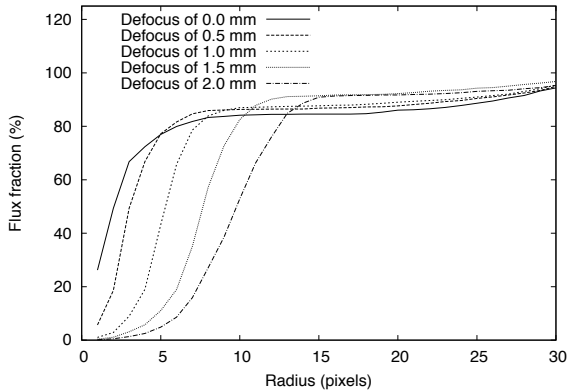}
\caption{Flux fraction within a given aperture versus aperture radius for 5 different levels of defocus, ranging from 0.0 to 2.0\,mm. These results were obtained for the simulated data, as described in the text.}  \label{fig:radProfileFF}
\end{figure}

To further investigate the effects of defocusing on photometry we simulated a star with integrated flux of $\sim10^6$ photons and a sky background flux of 28,000 photons/pixel. These values represent a typical WIRCam/CFHT exposure of a few seconds integration time, where the star flux is chosen to produce a maximum of about 23,000\,ADUs and the background is about 7,000\,ADUs. These values represent a compromise between exposure time and defocusing to get a reasonable amount of flux and to avoid the highly non-linear regime. Then we applied the measured PSF for each amount of telescope defocus to simulate the star data.  We performed aperture photometry for a number of aperture radii ranging from 1 to 30 pixels, where we calculated the signal-to-noise ratio (SNR) and the fraction of flux (FF) within the extraction aperture. \autoref{fig:radProfileSNR} shows a plot of SNR versus aperture radius and \autoref{fig:radProfileFF} shows FF versus aperture radius. The curves in \autoref{fig:radProfileSNR} show that focused images (defocus=0.0\,mm) reach the highest SNR very quickly, but the FF shown in \autoref{fig:radProfileFF} is not quite as large ($<70$\,\%) for the same aperture size (radius $< 3$\,pixels). A smaller photometric aperture decreases the total signal (by not including as much of the source flux), which may degrade the time-series photometry, since for smaller apertures, small offsets in the PSF will correspond to larger changes in the fraction of the total flux measured. On the other hand, by increasing the aperture size, the number of pixels and enclosed area of the sky also increases, thereby increasing the photon noise due to the background more rapidly than the signal from the source, and thereby decreasing the overall SNR. Therefore an optimal aperture for photometry may be used in order to obtain the best possible photometric precision.

\subsection{Detector characteristics}
\label{sec:detectorcharacteristics}

WIRCam is composed of a mosaic of 4 x 2048x2048 HAWAII-2RG detector arrays \citep{puget2004} with sky sampling of 0.301 arcsec/pixel, covering a total field-of-view of 21'.5 x 21'.5. WIRCam has 32 amplifiers per chip, where each amplifier is oriented along the horizontal (E-W) direction. The nominal value of pixel readout noise is $30\,e^{-}$ and the electronic gain is $3.8\,e^{-}/$ADU. Each amplifier presents slightly different sensitivities and noise characteristics, which cause spatial variations that may affect photometry.  The response of WIRCam pixels are not homogeneous and each pixel does not respond linearly to photon flux. This inhomogeneity can be partially corrected by standard flat-field correction, which is applied in all of our images. However, the non-linear response of the WIRCam array is more of an issue, which has been recently investigated and re-calibrated by the WIRCam team at CFHT (internal communications with Pascal Fouqu\'{e} and Wei-Hao Wang). We have been provided with these calibrations in two different flavors, where we adopted the standard non-linearity correction used at CFHT that applies a quadratic polynomial correction to each pixel. The correction considers a given pixel with raw measurement of $D$ counts in ADU, then the non-linearity correction ($NC$) is given by:

\begin{equation}
NC = a_{0} + a_{1} D + a_{2} D^{2},
\label{eq:nonlinquadraticcorr}
\end{equation}

where $a_{i}$ are the polynomial coefficients for the non-linearity correction. The corrected pixel value $D'$ is given by the raw value multiplied by $NC$, i.e., $D' = D\times NC$. The calibration provided by CFHT is given in a Multi-Extension FITS file containing 4 cube extensions (one for each WIRCam chip) with three slices in each cube (one for each coefficient). These calibrations also include bad pixel flags assigned to pixels where there was a failure in the fitting process to obtain the non-linearity correction function from calibrations, as reported by the CFHT team.

\section{Data Reduction and Analysis}
\label{sec:datareduction}

\subsection{Overview}

The reduction of our WIRCam data is performed by a custom pipeline. Our pipeline consists of a command line application written in Python that calls a series of libraries and modules written both in Python and C/C++. The C/C++ modules and libraries were adapted from the OPERA project \citep{martioli2012}. Our Python codes make use of the Astropy library \citep{astropy2013}, especially for FITS file handling, for astrometric tools, and for catalog query. In summary, our pipeline performs the following steps:

\begin{enumerate}
\item {\bf Pre-reduction calibrations}: flat-fielding, identification of targets in the 2MASS catalog \citep{skrutskie2006}, PSF measurements, image recentering, and aperture calibration;
\item {\bf Photometry}: flux extraction;
\item {\bf Differential photometry}: optimized selection of comparison stars based on magnitude, colour, field position, blending, and variability;
\item {\bf Analysis}: light curve detrending and eclipse model fitting.
\end{enumerate}

\subsection{Data Format}

WIRCam generates a data cube in FITS format for every sequence observed. The maximum number of slices supported for one sequence is 12. In all of our observations we used sequences of 12 exposures, where each slice in the cube contains 4 extensions, with one for each WIRCam chip. A typical observation of an eclipse event contains several dozen cubes. The reduction is performed in each WIRCam chip independently.  Each slice in the cube provides a photometric measurement for each source in the field-of-view.  We aimed to measure the integrated flux for each star in the field. Each star in all WIRCam chips was considered as a potential reference star to perform differential photometry on our target (\autoref{sec:cube}).   

\subsection{Flat-Field Calibration}

The first step in the reduction is the flat-field correction, where we divide each science frame by a normalized master flat-field. The master flat-field is calculated by the median of all individual normalized flat-field exposures. The flat-field exposures were obtained from sky observations during twilight, where we used the ones obtained in the nearest possible date to the night when science observations were done. The normalization of an individual flat-field exposure is done by dividing all pixel values by the median flux.

\subsection{Reference Cube Calibration}
\label{sec:cube}

In this step we take a reference cube to perform a number of calibrations. The reference cube selected is usually the first cube in the sequence. The pipeline generates a master reference image, by performing a median stack of all slices in the reference cube.   The master reference image is then used to perform the following calibration steps:

\begin{enumerate}

\item Query 2MASS catalog to generate a list of targets within the field-of-view observed. The sources are selected based on their K-magnitude. The maximum magnitude for source selection is set to be between 2 and 4 magnitudes above the target's 2MASS K-magnitude, depending on how crowded the field is. The main target is identified using a match with SIMBAD \citep{wenger2000}, therefore it must be identified by a known ID name on SIMBAD;

\item Perform basic astrometric calibration using the selected sources in the catalog. The astrometry performed in this step will fix WCS header keywords and will create a copy of the image extension with corrected astrometry. The astrometric correction is only applied to the zeroth order, precise to $\sim0.1$\,pixel, i.e.,  $\sim 0.03$ arcseconds;

\item Calibrate x,y positions of catalog sources using the master image.  The positions are first measured by the centroid of each star, and then an empirical Point Spread Function (PSF) is calculated from a median stack of the PSF from all selected stars in the field. Then, the star positions are recalculated through the maximum cross-correlation between the star flux and the measured PSF. We exclude targets that are either saturated or that were used for guiding. There is also a binarity check, where we exclude the faintest star in a pair which lies within a separation less than $50$\,pixels$\sim15$\,arcsec;

\item Perform basic photometry on the master reference image. This step also calculates a PSF which is obtained only from selected targets in the calibrated catalog. The new PSF is saved and used for further photometric calibration in the time series.  The PSF is recalculated for every cube and used to calibrate the center of the aperture used for source and sky flux extraction;

\item The last reduction step that is performed in the reference master image is the calibration of the aperture for photometry. We perform aperture photometry, where the flux is extracted within a circular aperture centered at the measured center of each star. A concentric annular aperture is used for sky flux measurements. As discussed in \autoref{sec:defocusandexptime}, both the signal-to-noise ratio (SNR) and the flux fraction (FF) of the source are a function of the aperture radius. We select an optimal aperture radius where both the SNR and FF calculated for a given incremental annulus become lower than a certain threshold. As an example, \autoref{fig:wasp-12-aperselect} shows the radial profile of $\sqrt{{\rm SNR_{\rm inc}*FF_{\rm inc}}}$ calculated for WASP-12, where we have also plotted a threshold of 2\%, providing an aperture radius of 16 pixels.

\begin{figure}
\includegraphics[width=\columnwidth]{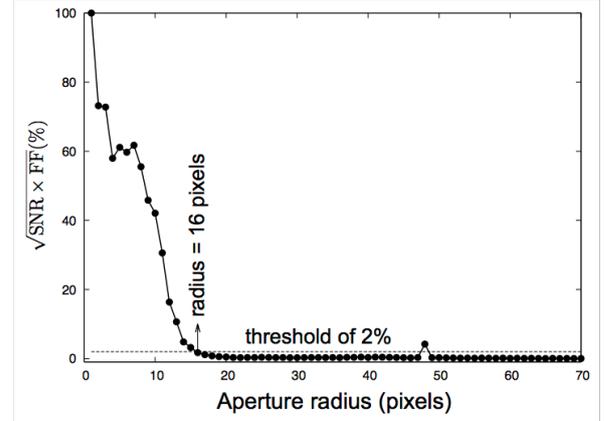}
\caption{This image illustrates the automatic selection of an aperture radius for photometry. Solid line shows the radial profile of the fractional increment of both the SNR and FF for target WASP-12. Dashed line shows the threshold of $\sqrt{{\rm SNR_{\rm inc}*FF_{\rm inc}}}<2\%$, which results in the selection of an aperture radius of 16 pixels.} \label{fig:wasp-12-aperselect}
\end{figure}

Now in order to test if this method is robust we calculate the photometric precision for the whole time-series for the eight brightest reference stars in the field of WASP-12. We perform aperture photometry using the same aperture size for all stars over the entire time series. The photometry was repeated for several aperture sizes with radius ranging from 14 to 24 pixels. The photometric precision is defined here as the mean standard deviation, where each individual standard deviation is calculated as the standard deviation around the mean of a short sequence of 12 images. The results shown in \autoref{fig:aper-phot-precision} indicate that the best photometric precision is attained for an aperture with radius around 16 pixels, which is consistent with the size obtained by using a threshold of $\sqrt{{\rm SNR_{\rm inc}*FF_{\rm inc}}}$ $<$ 2\% as presented in \autoref{fig:wasp-12-aperselect}.

\begin{figure}
\includegraphics[width=\columnwidth]{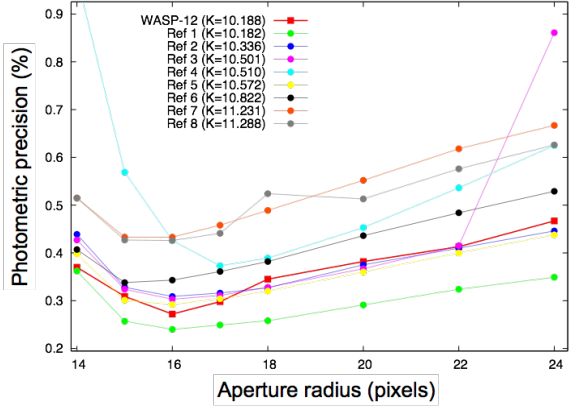}
\caption{Photometric precision versus aperture radius for the eight brightest reference stars in the field of WASP-12. The K-magnitude of each star is given in the legend.}  \label{fig:aper-phot-precision}
\end{figure}

\end{enumerate}

\subsection{Photometry}

The PSF is measured in every cube of the time series, where the PSF is used for re-centering the aperture position and for extracting the flux of all frames within the respective cube length. The photometric measurements are performed on each source independently.  The following algorithm is used for each individual photometric measurement.

\begin{figure}
\includegraphics[width=\columnwidth]{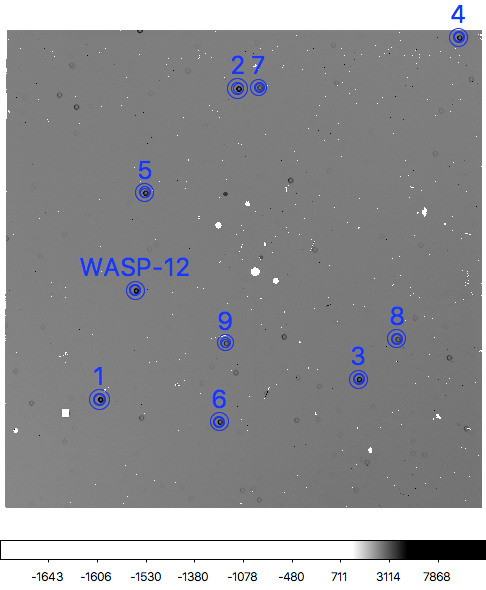} 
\caption{WIRCam exposure of WASP-12, showing examples of the apertures for photometry for the ten brightest stars in the field. The inner circle in each aperture is the region used for object flux measurements and the outer annulus is the region used for sky flux measurements.} \label{fig:apertureExample}
\end{figure}

\begin{enumerate}

\item Recenter star position to the maximum cross-correlation between the star flux and the measured PSF. Then, the aperture center is reset to the updated star position;
\item Sky flux measurements. The sky flux $S$ is measured on an annular aperture. The sky aperture inner radius is offset by about 8 pixels from the main aperture and the outer radius is calculated being two times the radius of the main aperture. If the radial distance between the inner and outer radii is smaller than 2 pixels then the annulus size is increased to have a width of 2 pixels, so it will cover a larger area in the sky. The sky flux is calculated as the median of all individual measurements within the annulus, i.e., 

\begin{equation}
S={\rm MEDIAN}(s_{i}), 
\label{eq:skyflux}
\end{equation}

where the sky flux $s_{i}$ on each pixel $i$ is calculated as follows:

\begin{equation}
s_{i} = d_{i}\times NC \times G,
\label{eq:pixelflux}
\end{equation}

for $d_{i}$ being the pixel value in ADU, $NC$ being the non-linearity correction calculated by \autoref{eq:nonlinquadraticcorr}, and $G$ the gain in e-/ADU;

\item Source flux measurements. The source flux is calculated as the integrated flux within a circular aperture. The aperture size is kept constant in the entire time series. \autoref{fig:apertureExample} shows an example of a WIRCam exposure of WASP-12 field, where we also show the apertures (blue circles) used for photometry. 

The source flux $F$ is calculated as the sum of all sky-subtracted fluxes of useful pixels within the aperture, which is given by the following equation:

\begin{equation}
F = \frac{\sum_{i} {M_{i}\times \left(d_{i}\times NC \times G - S \right)}}{\sum_{i} {M_{i}\times\phi_{i}}},
\label{eq:sourceflux}
\end{equation}

where $S$ is the sky flux given by \autoref{eq:skyflux}, the factor $M_{i}$ in \autoref{eq:sourceflux} is a Boolean mask function to avoid the contribution from bad pixels (non-useful pixels), where $M_{i}=0$ when the pixel is identified either as a bad pixel or as a missing pixel and $M_{i}=1$ if the pixel is considered good. The function $\phi_{i}$ is the expected source flux fraction for a given pixel $i$, which is directly obtained from the measured PSF.

The variance $\sigma_{F}^{2}$ in each flux measurement is assumed to be the sum of Poisson photon noise, electronic readout noise, and sky flux variance. Therefore the flux variance is calculated as follows:

\begin{equation}
\sigma_{F}^{2} = \frac{\sum_{i} {M_{i}\times \left[(d_{i}\times NC \times G - S) + N^{2} + \sigma_{S}^{2} \right]}}{\sum_{i} {M_{i}\times\phi_{i}}},
\label{eq:sourcefluxVar}
\end{equation}

where $N$ is the readout noise, and $\sigma_{S}$ is the median deviation in the sky flux measurements. 

\end{enumerate}

\subsection{Differential Photometry}
\label{sec:diffphot} 

The product from the previous step is a vector of flux measurements and variances for each selected source in the catalog. Each of these vectors is a raw light curve, which is ingested by the pipeline to perform the differential photometry analysis. The light curve analysis consists basically of the following steps:

\begin{enumerate}

\item {\bf Merge light curves}. As mentioned earlier, each WIRCam detector chip is treated separately in the reduction. However, only one chip contains the main target. Therefore, at this point the pipeline will merge all light curves into a single table containing all stars observed, so one can perform differential analysis using comparison stars observed on other chips as well.

\item {\bf Differential photometry}. In this step the pipeline calculates the ratio between the main target flux $F_{T}$ and the flux of each comparison star $F_{j}$, i.e., ${\rm R}_{j} = F_{T}/F_{j}$, where $j=1,...,n_{c}$, and $n_{c}$ is the total number of comparison stars. The variance of each relative measurement is given by $\sigma_{{\rm R}_j}^{2} = {\rm R}^{2}_{j}  (\sigma_{T}^{2}/F_{T}^{2} + \sigma_{j}^{2}/F_{j}^{2})$, where we assume the errors from different stars are not correlated.  

\item \label{item:detrendingfirstpass} {\bf Detrending (first pass)}. The light curve of each comparison star is first analyzed separately.  A linear function $L_{j}(t)$ is fit to the flux ratio data of each comparison star, where we use the robust LADFIT \citep[Least Absolute Deviation;][]{bloomfieldandsteiger1983} method for the fit.  Then each light curve is normalized by the linear trend, i.e., the detrended light curve ${\rm R'}_{j}$ is given by ${\rm R'}_{j} = {\rm R}_{j} / L_{j}$.

\item {\bf Selection of comparison stars}. One main criterion is used for selecting comparison stars. The pipeline calculates the photometric precision for each comparison star through the standard deviation of the detrended individual light curves obtained in \autoref{item:detrendingfirstpass}. Then it selects targets for which the photometric precision is lower than a given threshold. This threshold is typically chosen between 0.5\% and 1\%. However if plenty of bright stars are available one should set it to lower values to reach the best possible photometric precision, otherwise the flux from faint stars may degrade the precision of the final light curve. Even though the pipeline frequently provides a good set of comparison stars, before getting the final light curve, we analyze each light curve individually by eye, where we may  exclude those stars presenting variability or those highly correlated with the airmass of observations. 

\item {\bf Simultaneous detrending and eclipse model}. Once a suitable and clean set of comparison stars is selected, the pipeline performs a simultaneous fit to all differential light curves, where the model includes both a background trend and the eclipse model. With this approach the background is modeled by a quadratic polynomial, $B_{j}(t) = a_{j} t^{2} + b_{j} t + c_{j}$, where the coefficients $a_{j}$, $b_{j}$, and $c_{j}$ are independent for each comparison star. The parameters in the eclipse model are constrained to be equal for all comparison stars. The fit procedure using Bayesian inference is presented in more detail in \autoref{sec:lightcurvemodel}.

\item {\bf Final light curve}. Once the background ``trends" are obtained using the method explained above, one can reduce the light curves by dividing each individual flux ratio by the polynomial $B_{j}(t)$. Then all light curves are combined into a final light curve by calculating either the mean or the median of all individual detrended flux ratios, i.e., ${\rm R}_{T} = {\rm MEAN}\left(R'_{j}, i=1,..,n_{c} \right)$ or ${\rm R}_{T} = {\rm MEDIAN}\left( R'_{j}, i=1,..,n_{c} \right)$. The ``MEAN'' is preferred, but only when the light curves do not present a large amount of outliers. 

\item {\bf Binning}. Finally, the pipeline bins the data by the median of points within a given time bin. We typically use a bin size with the length of 2 cubes, i.e., bin size of 24 data points. The bin size is not relevant in our analysis, since we always use the original data to perform the fit, but it is useful to produce a better visualization of the uncertainties.

\end{enumerate}

\autoref{table:reductionlog} presents the optimal aperture sizes calculated for each target in our sample. It also presents the number of reference stars, the WIRCam chips where these stars were located, the photometric precision cutoff ($\sigma_{max}$) used to select reference stars, and the final RMS of residuals calculated after both the trends and the eclipse model have been removed.

\begin{table}
\caption{Log of reduction.}
\label{table:reductionlog}
\begin{tabular}{cccccc}
\hline
Object ID & Aperture & \# Ref. & Chips & $\sigma_{max}$ & $\sigma_{\rm res}$ \\
& [pixels] &  Stars  &  & [\%] & [\%]\\
\hline
\hline
WASP-12 & 16 & 4 & 0,1 &  0.32 &  0.20 \\
KELT-4A & 17 & 1 & 0 & 0.28 & 0.28 \\
WASP-14 & 17 & 4 & 2,3 & 0.44 &  0.24 \\
TrES-4 & 15 &  3 & 0,2 & 0.70 & 0.53 \\
Kepler-5 & 15 &  6 & 0 &  0.45 & 0.38 \\
KELT-2A & 16 & 4 & 0,1,2,3 & 0.38 &  0.22 \\
KELT-7 & 14 &  2 & 0&  0.17 & 0.11 \\
WASP-31 & 14 & 2 & 0 &  0.34 & 0.24 \\
HAT-P-33 & 14 & 3 & 0,2 & 0.36 & 0.19 \\
\hline
\end{tabular}
\end{table}

\subsection{Light curve model}
\label{sec:lightcurvemodel}

We implemented a parameter estimation Bayesian analysis to model both the background ``trends" and the eclipse simultaneously. The eclipse model is calculated using the BATMAN package by \citet{kreidberg2015}, which implements the \citet{mandelagol2002} transit model for eclipses. Since we are dealing with eclipses and not transits, we removed the contribution from stellar limb-darkening. Assuming this model is adequate to describe our observations, we implemented a Bayesian posterior probability estimation analysis, where we applied the \citet{goodmanweare2010} affine invariant Markov chain Monte Carlo (MCMC) ensemble sampler using the emcee package by \citet{foreman-mackey2013}. We followed the steps described in the online tutorial \footnote{\label{foot:parviainentutorial} \url{http://dan.iel.fm/emcee}} to infer the posterior probability for the eclipse model parameters and the background polynomial coefficients simultaneously. The priors from which the eclipse parameters are sampled were obtained from prior knowledge of the systems, e.g. from transit predictions. We have used the parameters and uncertainties presented in \autoref{table:parspriors}. In order to optimize the fit process, each parameter have been assigned a different prior probability distribution. For the orbital period ($P$), the planet radius ($r_{p}$), and semi-major axis ($a$), we adopted a normal distribution. The orbital inclination ($i$), eccentricity ($e$), and longitude of periastron ($\omega$), are fixed as constants. For the planet-to-star flux ratio ($\delta$) we adopted an uniform distribution between $0\,\% < \delta < 0.5\,\%$. The central time of eclipse ($t_{sec}$) is also sampled from an uniform distribution between $-0.020\,d < t_{sec} - t_{c} < 0.020\,d$, where $t_{c}$ is the predicted central time of eclipse.

\begin{table*}
\caption{The eclipse parameters priors. For $t_{sec}$ and $\delta$ the prior probability distribution (PPD) is uniform, where we adopted U($t_{c}-0.020$,$t_{c}+0.020$) and U(0,0.5), respectively.  For $P$, $r_{p}$, and $a$ the PPD is normal, i.e., N($\mu,\sigma$), where $\mu$ is the central value and $\sigma$ is the error. For $i$, $e$, and $\omega$ the prior is a constant value.}
\label{table:parspriors}
\begin{tabular}{lccccccccc}
\hline
Object ID & $t_{c}$\,(MJD) & $\delta$\,(\%) & $P$\,(d) & $r_{p}/R{*}$& $a/R{*}$ & $i\,(^{\circ})$ & $e$ & $\omega\,(^{\circ})$ & Reference\\	  
\hline
\hline
WASP-12  & $55194.434$ & U(0,0.5) & $1.091422\pm1\times10^{-6}$ & $0.109\pm0.008$ & $3.1\pm0.2$ & $86.0$ & $0.0$ & $90$ & \citet{chan2011} \\
KELT-4A  & $56735.382$ & U(0,0.5) & $2.989593\pm5\times10^{-6}$ & $0.106\pm0.007$ & $5.8\pm0.3$ & $83.1$ & $0.03$ & $300$ & \citet{eastman2016} \\
WASP-14  & $56795.457$  & U(0,0.5) & $2.243766\pm1\times10^{-6}$ & $0.099\pm0.008$ & $5.9\pm0.4$ & $84.8$ & $0.087$ & $252.9$ & \citet{blecic2013}\\
TrES-4   & $56851.437$  & U(0,0.5) & $3.553927\pm3\times10^{-6}$ & $0.095\pm0.004$ & $6.1\pm0.2$ & $82.8$ & $0.0$ & $90$ & \citet{chan2011} \\
Kepler-5 & $56873.344$  & U(0,0.5) & $3.54846\pm3\times10^{-5}$ & $0.080\pm0.004$ & $6.1\pm0.2$ & $86.3$ & $0.0$ & $90$ & \citet{borucki2010} \\
KELT-2A  & $57000.489$  & U(0,0.5) & $4.11379\pm1\times10^{-5}$ & $0.071\pm0.005$ & $6.4\pm0.3$ & $90.0$ & $0.185$ & $160$ &  \citet{beatty2012} \\
KELT-7   & $57001.495$  & U(0,0.5) & $2.734775\pm4\times10^{-6}$ & $0.089\pm0.004$ & $5.5\pm0.2$ & $83.8$ & $0.0$ & $90$ & \citet{bieryla2015} \\
WASP-31  & $57053.514$  & U(0,0.5) & $3.405909\pm5\times10^{-6}$ & $0.125\pm0.006$ & $8.1\pm0.3$ & $84.5$ & $0.0$ & $90$ & \citet{anderson2011} \\
HAT-P-33 & $57054.389$  & U(0,0.5) & $3.474474\pm1\times10^{-6}$ & $0.103\pm0.023$ & $6.1\pm1.0$ & $86.7$ & $0.148$ & $96$ & \citet{hartman2011} \\
\hline
\end{tabular}
\end{table*}

\begin{figure*}
\includegraphics[scale=0.6]{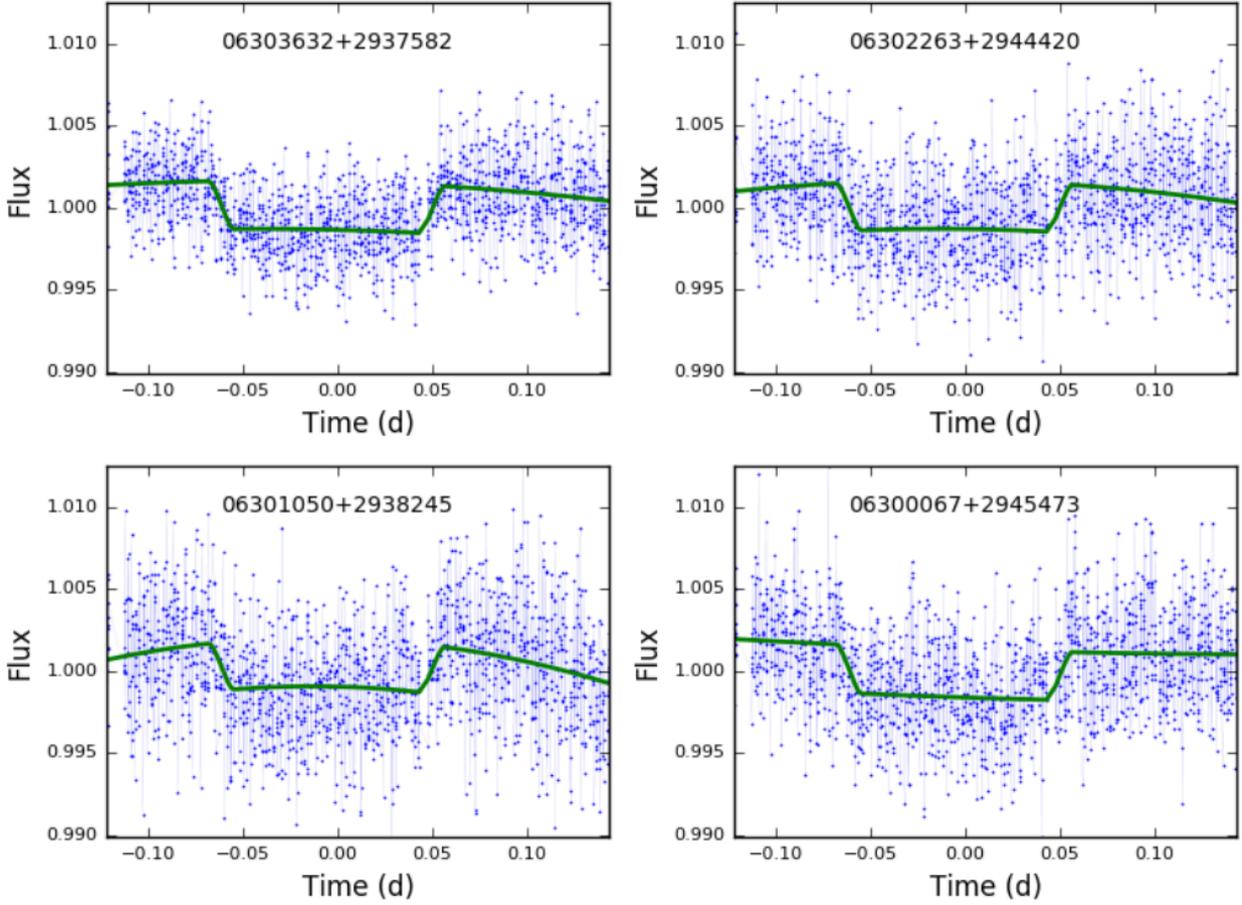} 
\caption{WASP-12 differential photometry light curves for all selected comparison stars. Each light curve is identified by the 2MASS ID of the comparison star. The solid green lines show the global fit model, which includes the background trends and the eclipse model, as explained in the text.} \label{fig:WASP-12redphot}
\end{figure*}

\begin{figure*}
\includegraphics[scale=0.5]{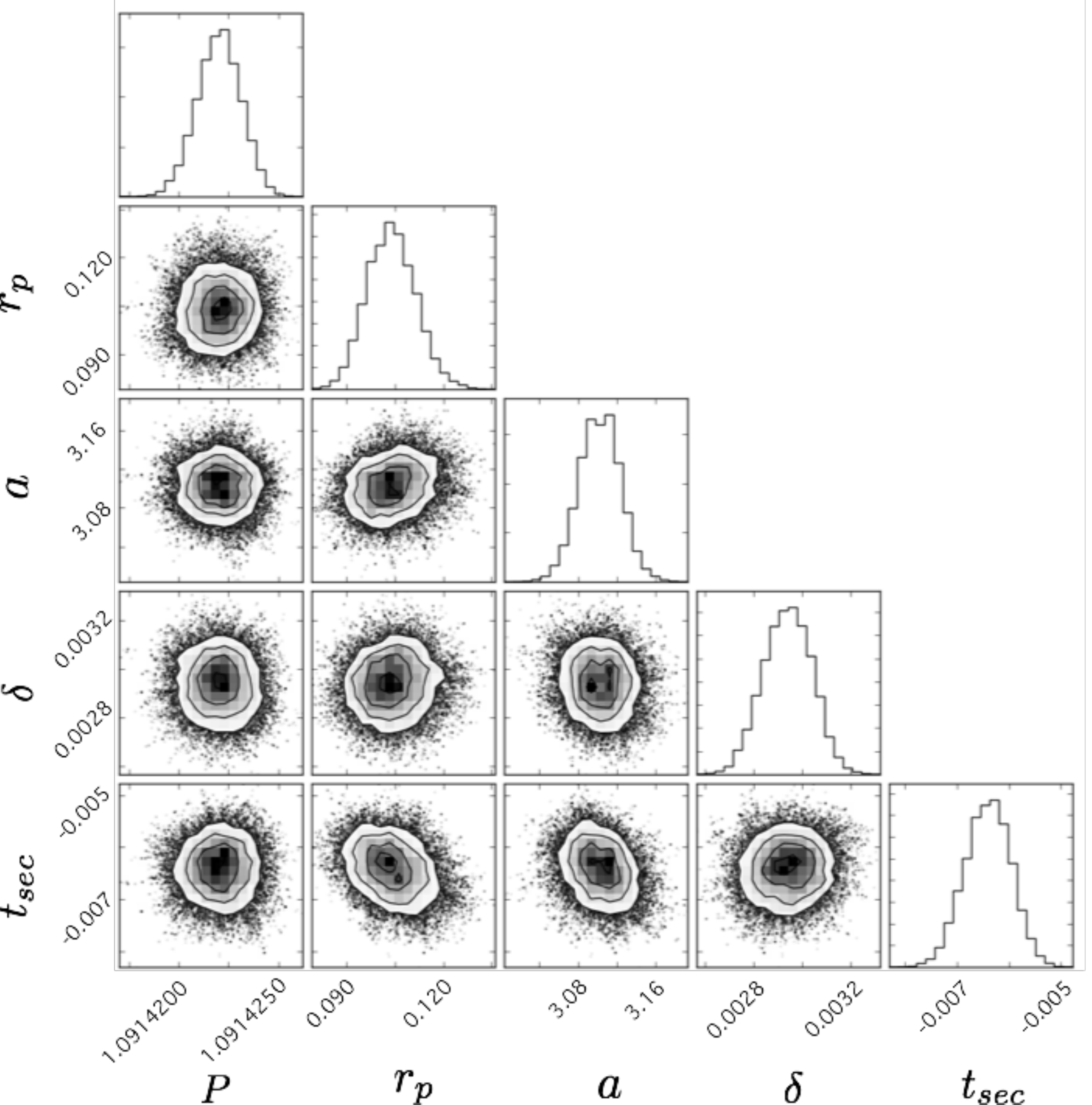} 
\caption{Corner plot showing the one and two dimensional projections of the posterior probability distributions for the five eclipse parameters ($t_{sec}, \delta, P, r_{p}, a$). The fit is made simultaneously for all model parameters, i.e., including the three background coefficients for each comparison star (total of 12), however we only plotted the eclipse parameters for the sake of clarity. } \label{fig:WASP-12pairs}
\end{figure*}

In \autoref{fig:WASP-12redphot} we present the differential photometry light curves for WASP-12 with respect to the four stars selected as comparison.  We have run our analysis on these data and obtained the model represented by the green line in \autoref{fig:WASP-12redphot}. Notice the baseline of each individual light curve can be reliably modeled by a quadratic polynomial function to account for systematics. \autoref{fig:WASP-12pairs} shows the one and two dimensional projections of the posterior probability distributions obtained for all the background and eclipse parameters. These distributions were obtained by running 3000 iterations of the MCMC sampler, where we discarded the first 1000 samples as burn-in.

\begin{figure}
\includegraphics[scale=0.44]{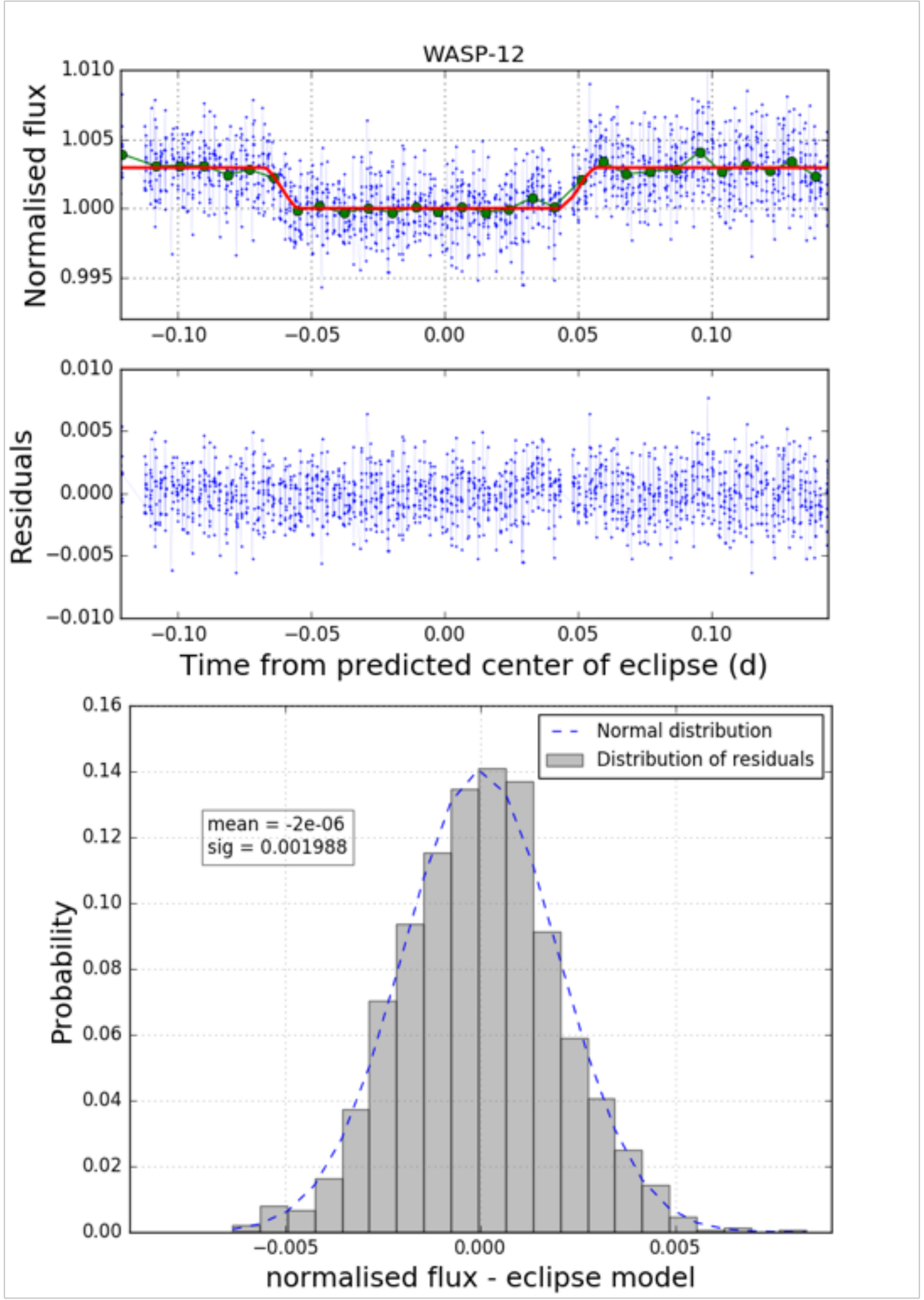} 
\caption{Top panel shows the reduced light curve (blue circles connected by thin line) for the control target WASP-12b, the binned data (green circles), and the eclipse model (red solid line). Middle panel shows the residuals. Bottom panel shows the probability distribution function of residuals (gray bars) and a normal distribution model (blued dashed line), which is calculated using the mean ($\mu$) and standard deviation ($\sigma$) of residuals. The values of $\mu$ and $\sigma$ are also presented.} \label{fig:wasp-12}
\end{figure}

\autoref{fig:wasp-12} shows the final reduced light curve and the eclipse fit model for the control target WASP-12b. \citet{croll2015} obtained flux ratio of $\delta=0.284^{+0.019}_{-0.020}$\%, which is in good agreement with our measured flux ratio of $\delta=0.294\pm0.010$\%.  In \autoref{fig:wasp-12} we also present the probability distribution for the residuals and a normal distribution model N($\mu$,$\sigma$) for comparison, where $\mu$ and $\sigma$ are the mean and standard deviation of residuals, respectively. For our control target WASP-12b, the residuals seem to be normally distributed, since the normal model calculated from $\mu$ and $\sigma$ is in good agreement with the probability distribution.

\begin{figure}
\includegraphics[width=\columnwidth]{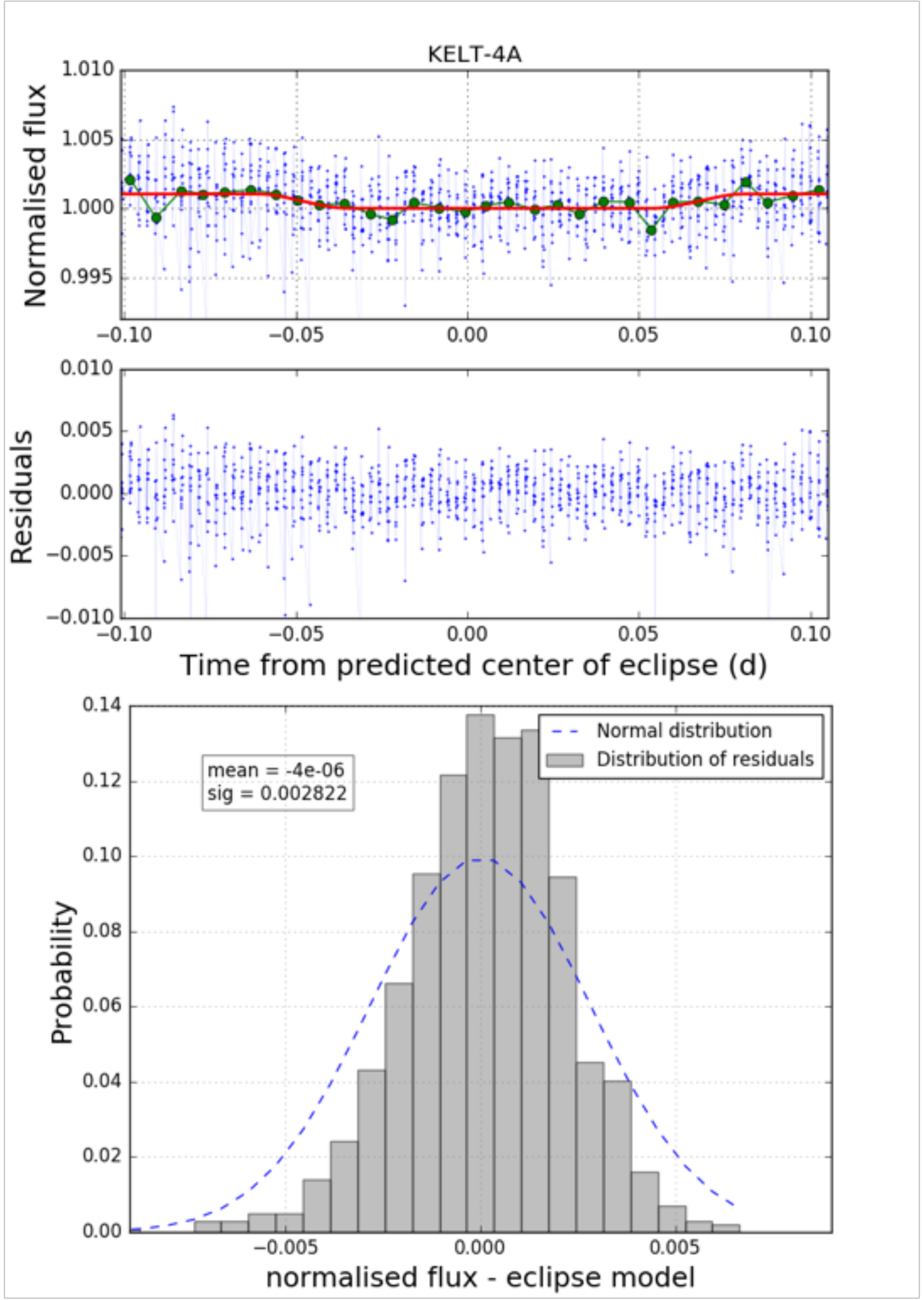} 
\caption{Same as \autoref{fig:wasp-12} for target KELT-4A.} \label{fig:KELT-4}
\end{figure}

\begin{figure}
\includegraphics[width=\columnwidth]{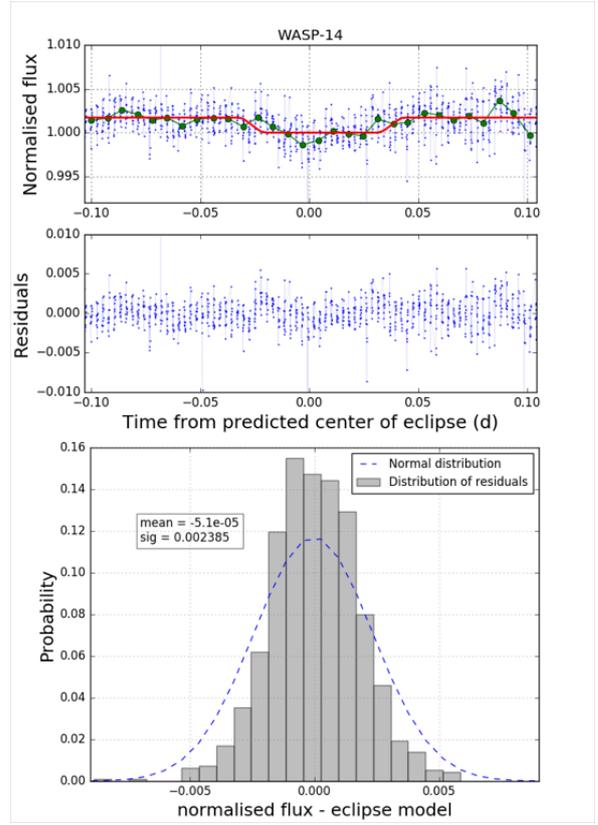} 
\caption{Same as \autoref{fig:wasp-12} for target WASP-14.} \label{fig:WASP-14}
\end{figure}

\begin{figure}
\includegraphics[width=\columnwidth]{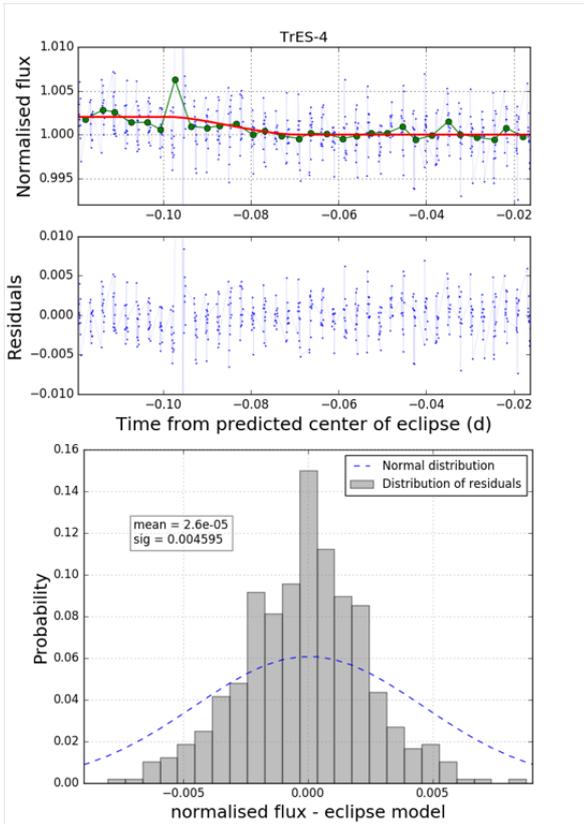} 
\caption{Same as \autoref{fig:wasp-12} for target TrES-4.} \label{fig:TrES-4}
\end{figure}

\begin{figure}
\includegraphics[width=\columnwidth]{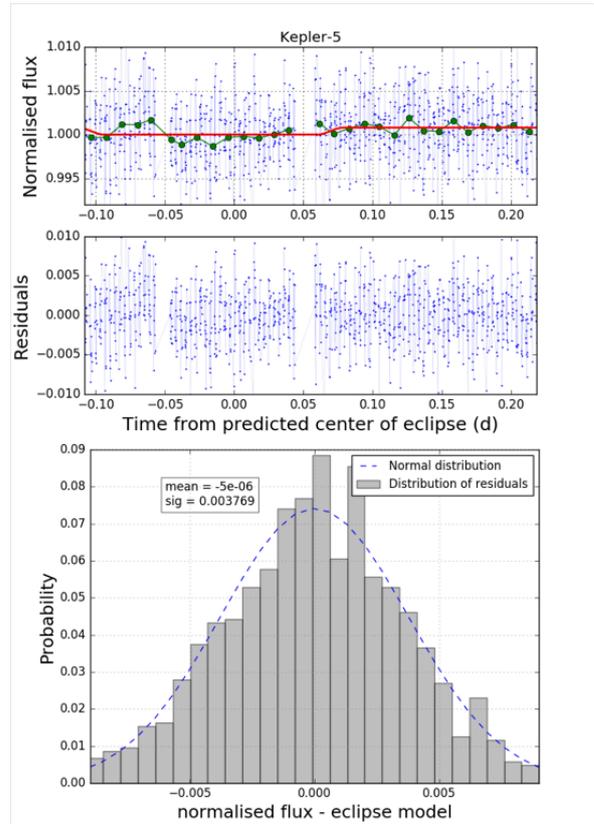} 
\caption{Same as \autoref{fig:wasp-12} for target Kepler-5.} \label{fig:Kepler-5}
\end{figure}

\begin{figure}
\includegraphics[width=\columnwidth]{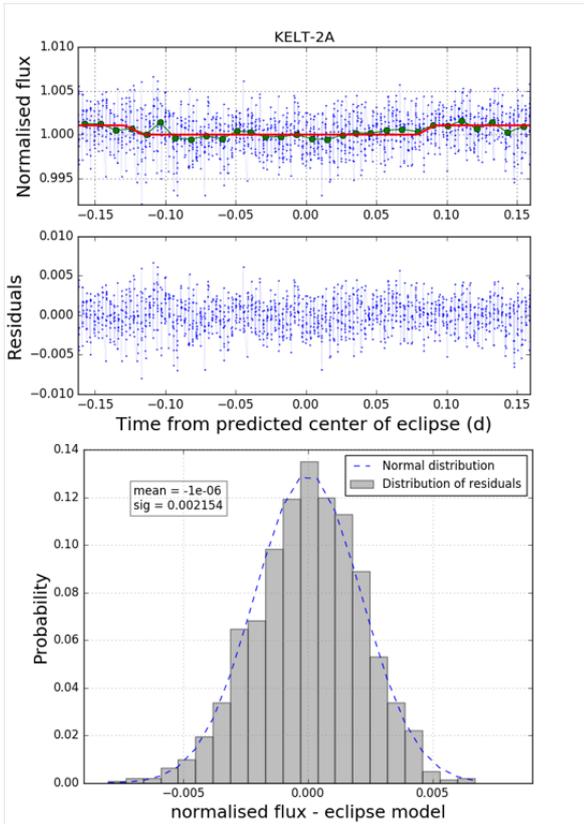} 
\caption{Same as \autoref{fig:wasp-12} for target KELT-2A.} \label{fig:KELT-2A}
\end{figure}

\begin{figure}
\includegraphics[width=\columnwidth]{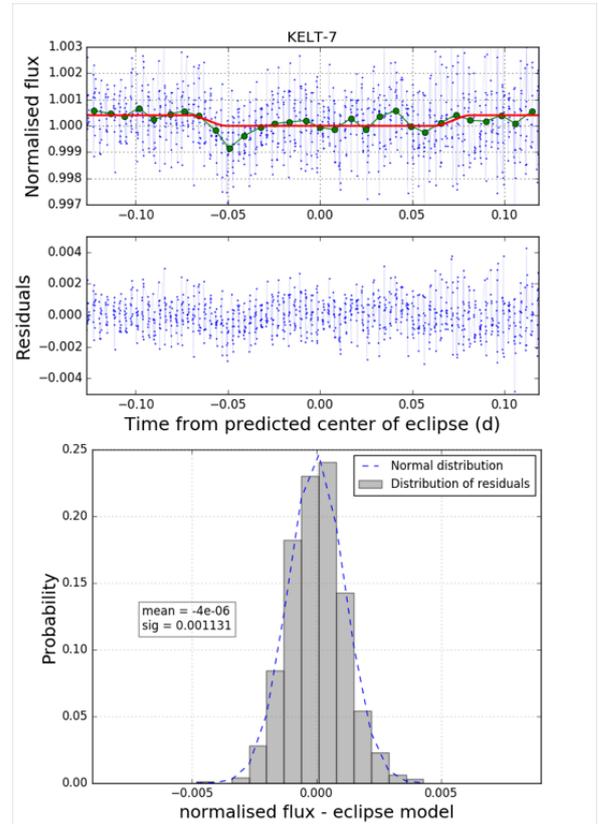} 
\caption{Same as \autoref{fig:wasp-12} for target KELT-7.} \label{fig:KELT-7}
\end{figure}

\begin{figure}
\includegraphics[width=\columnwidth]{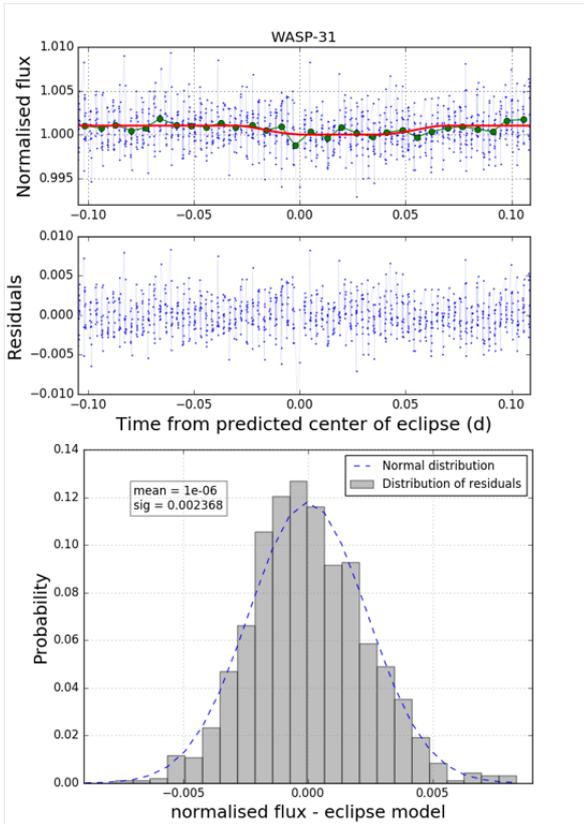} 
\caption{Same as \autoref{fig:wasp-12} for target WASP-31.} \label{fig:WASP-31}
\end{figure}

\begin{figure}
\includegraphics[width=\columnwidth]{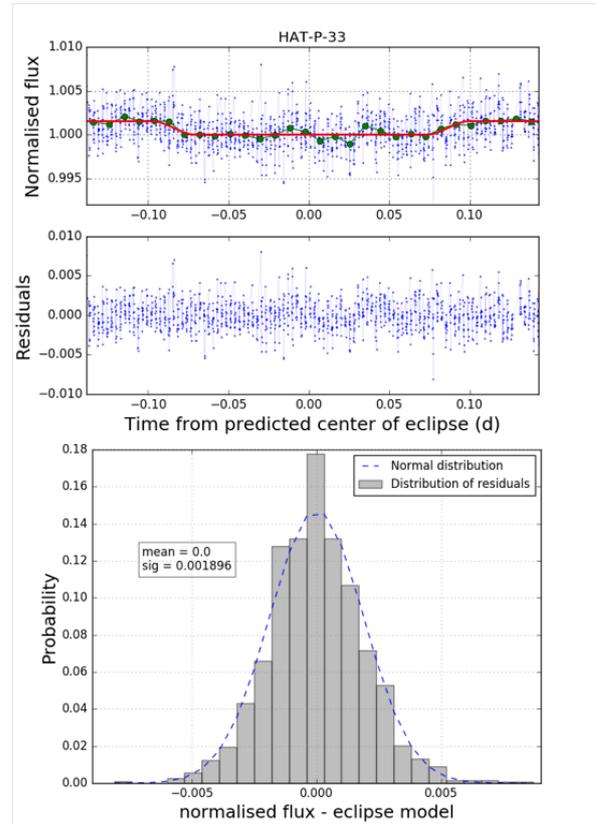} 
\caption{Same as \autoref{fig:wasp-12} for target HAT-P-33.} \label{fig:HAT-P-33}
\end{figure}

Similarly, \autoref{fig:KELT-4}, \autoref{fig:WASP-14}, \autoref{fig:TrES-4}, \autoref{fig:Kepler-5}, \autoref{fig:KELT-2A}, \autoref{fig:KELT-7}, \autoref{fig:WASP-31}, and \autoref{fig:HAT-P-33} show the light curves, fit models, and the analysis of residuals for all objects in our sample. \autoref{table:fitpars} shows all the final fit parameters and uncertainties derived from the posterior distribution. Notice that the light curves for several objects in our sample still present some residual red noise (e.g. WASP-14, TrES-4, Kepler-5, KELT-7), which is not accounted for in our analysis.  Therefore the uncertainties in the eclipse depths may be underestimated for some of our targets.

\begin{table*}
\caption{Fit parameters and uncertainties. The central value of each parameter is the 50th percentile of the posterior distribution, and the uncertainty is calculated by the average between the 50th minus 16th percentiles, and the 84th minus 50th percentiles.}
\label{table:fitpars}
\begin{tabular}{lccccc}
\hline
Object ID & $t_{sec}$ $^{b}$ & $\delta$\,(\%) & $P$\,(d) & $r_{p}/R{*}$ & $a/R{*}$ \\	  
\hline
\hline
WASP-12  & $55194.4276\pm0.0004$ & $0.294\pm0.010$ & $1.091422\pm1\times10^{-6}$ & $0.1038\pm0.0075$ & $3.103\pm0.021$ \\
KELT-4A  & $56735.3940\pm0.0029$ & $0.172\pm0.029$ & $2.989593\pm5\times10^{-6}$ & $0.1069\pm0.0069$ & $5.874\pm0.131$ \\
WASP-14  & $56795.4632\pm0.0014$ & $0.172\pm0.025$ & $2.243766\pm1\times10^{-6}$ & $0.1005\pm0.0093$ & $7.820\pm1.159$ \\
TrES-4   & $56851.4200\pm0.0040$ & $0.202\pm0.090$ & $3.553927\pm3\times10^{-6}$ & $0.0945\pm0.0039$ & $5.882\pm0.140$ \\
Kepler-5 & $56873.3259\pm0.0023$ & $0.080\pm0.025$ & $3.548457\pm3\times10^{-6}$ & $0.0795\pm0.0041$ & $6.047\pm0.039$ \\
KELT-2A  & $57000.4701\pm0.0010$ & $0.105\pm0.015$ & $4.113789\pm9\times10^{-6}$ & $0.0702\pm0.0050$ & $6.631\pm0.197$ \\
KELT-7   & $57001.4984\pm0.0048$ & $0.040\pm0.012$ & $2.734775\pm4\times10^{-6}$ & $0.0890\pm0.0040$ & $5.382\pm0.215$ \\
WASP-31  & $57053.5335\pm0.0006$ & $0.102\pm0.017$ & $3.405909\pm5\times10^{-6}$ & $0.1262\pm0.0059$ & $8.644\pm0.197$ \\
HAT-P-33 & $57054.3900\pm0.0016$ & $0.153\pm0.022$ & $3.474474\pm1\times10^{-6}$ & $0.1085\pm0.0198$ & $6.741\pm0.105$ \\
\hline
\multicolumn{6}{l}{\footnotesize{$^{b}$ Eclipse centre times are in BJD$-2450000$}}

\end{tabular}
\end{table*}
 
\subsection{Dilution correction}
\label{sec:dilutioncorrection}

Our analysis so far presents the flux ratios obtained from the data as is. However, some of our targets are known to have nearby companion stars blended with them, i.e., lying within the $\sim5$ arcsec apertures used in our photometry.  Therefore the flux measurements for those blends need to be corrected. We correct the measured depth ($\delta$) by applying a dilution factor $\epsilon$, as in \citet{zhao2014}. Thus the final undiluted planet-to-star flux ratio is given by $f_{p}/f_{A}=\epsilon \delta$, where

\begin{equation}
\epsilon = \left(1 + \frac{f_{B}}{f_{A}} \right),
\label{eq:dilutionfactor}
\end{equation}

where $f_{B}/f_{A}$ is the flux ratio between the sum of fluxes of all other components in the system (B,C, etc.) and the planet host component A. \autoref{table:dilutioncorrection} shows the differential magnitudes in K-band ($\Delta K$) between component A and other known components. It also shows the derived dilution factor ($\epsilon$), and the dilution-corrected occultation depths ($f_{p}/f_{A}$) for the three systems in our sample with known nearby stellar companions: WASP-12 \citep{bechter2014}, KELT-4 \citep{eastman2016}, and KELT-2 \citep{beatty2012}. Note that for the system KELT-2 we have not found available K-magnitudes for each individual star in the system. Therefore, we calculated $\Delta K$ based on the V-magnitudes and the system's K-magnitude given in \citet{beatty2012}, and we converted between V to K magnitudes using the intrinsic colour of dwarfs (K-V) by \citet{besselbrett1988}.

\begin{table*}
\caption{Dilution correction.}
\label{table:dilutioncorrection}
\begin{tabular}{cccccc}
\hline
System ID & Components & $\Delta K$ & $\epsilon$ & $f_{p}/f_{A}$ (\%) & Ref. \\
\hline
\hline
WASP-12 & A,B,C & $2.51\pm0.03\,^{a}$ & $1.0991\pm0.0027$ & $0.323\pm0.011$ & \citet{bechter2014} \\
KELT-4  & A,B,C & $1.38\pm0.14\,^{a}$ & $1.28\pm0.04$     & $0.220\pm0.040$   & \citet{eastman2016} \\
KELT-2  & A,B   & $2.25\pm0.20\,^{b}$ & $1.126\pm0.023$   & $0.118\pm0.017$ & \citet{beatty2012}  \\
\hline
\multicolumn{6}{l}{\footnotesize{$^{a}$ $\Delta K = K_{\rm B+C} - K_{\rm A}$}} \\
\multicolumn{6}{l}{\footnotesize{$^{b}$ $\Delta K =  K_{\rm B} - K_{\rm A}$}}
\end{tabular}
\end{table*}

\section{Discussion}
\label{sec:discussion}

\begin{figure}
\includegraphics[width=\columnwidth]{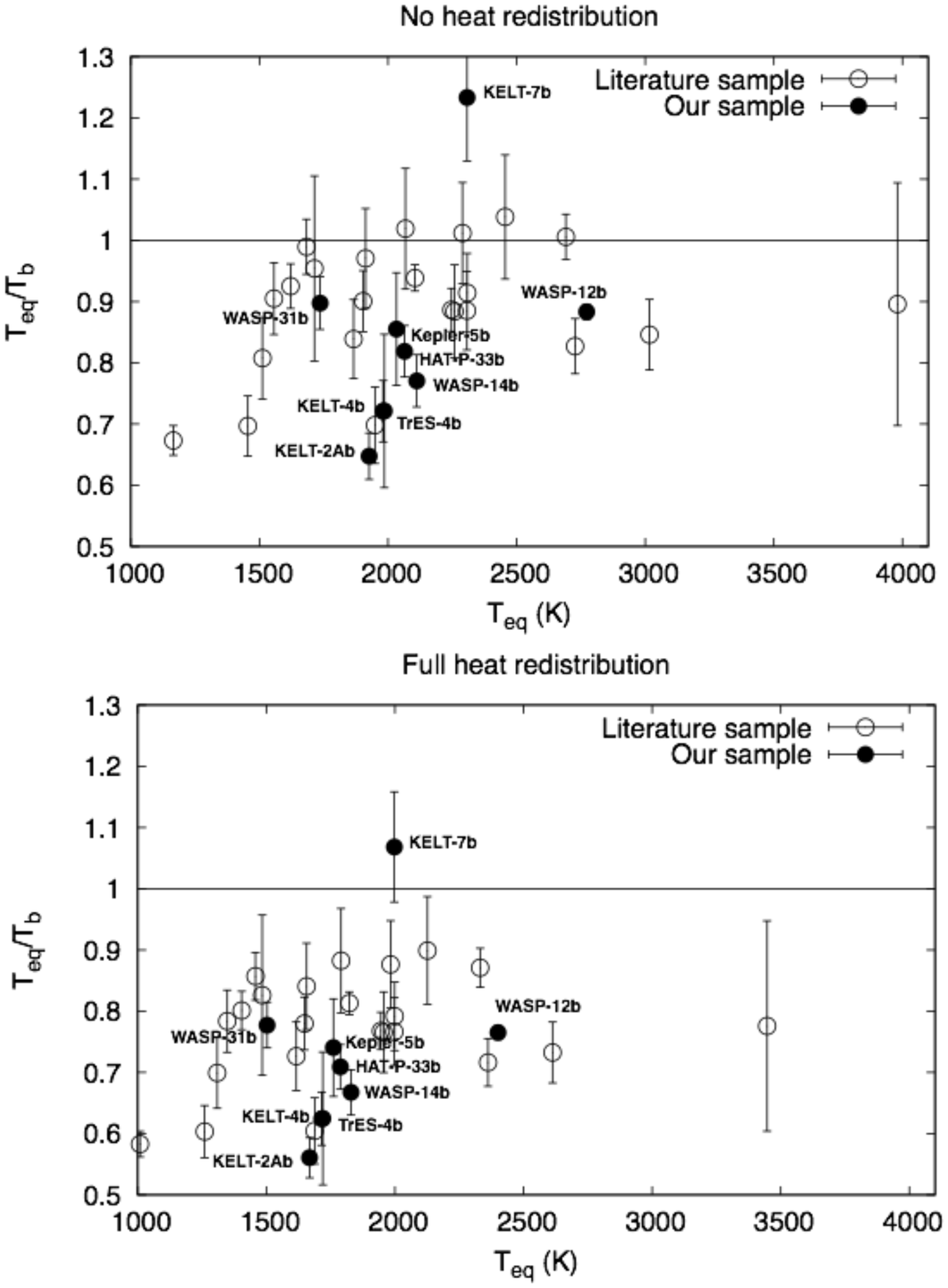} 
\caption{Equilibrium temperature (T$_{\rm eq}$) versus the ratio between equilibrium temperature and the brightness temperature (T$_{\rm eq}$/T$_{\rm b}$), where T$_{\rm b}$ was obtained from the measured flux ratio in K-band assuming both the planet and the star emit as blackbodies, and T$_{\rm eq}$ was calculated assuming bond albedo of 0.1 and heat redistribution factor for both no heat redistribution $f=2/3$ (top panel) and full heat redistribution $f=1/2$ (bottom panel). Filled circles show our data and open circles show the data from the literature as presented in \autoref{table:exosystems}. The error bars were calculated considering only the error in the flux ratio.} \label{fig:TeqXTeqOverTb}
\end{figure}

\autoref{table:exosystems} shows the exoplanet systems with known planet-to-star flux ratios measured in the K-band. These include the literature sample collated in Table A1 of \citet{zhou2015} and also the systems with detected secondary eclipses reported in this paper. We adopted the system's parameters shown in \autoref{table:exosystems} to calculate the equilibrium temperature $T_{\rm eq}$ of the exoplanet as given in \citet{henganddemory2013}, i.e.,

\begin{equation}
T_{\rm eq} = T_{*} \left( \frac{R_{*} f}{a} \right)^{1/2} (1 - A)^{1/4},
\label{eq:equilibriumtemperature}
\end{equation}

where we assume an albedo of $A=0.1$, which is consistent with Rayleigh scattering caused by hydrogen molecules alone \citep{sudarsky2000}, and we also assume both an uniform heat redistribution ($f=1/2$) and no heat redistribution ($f=2/3$), i.e., no heat is transported from the dayside to the nightside of the planet. We also calculated the brightness temperature $T_{\rm b}$ as defined in \citet{seageranddeming2010}, where we used the measured planet-to-star flux ratio in the K-band ($\lambda \sim2.2\mu$\,m) and assumed that both the star and the planet emit as blackbodies. \autoref{fig:TeqXTeqOverTb} presents a plot of the ratio $T_{\rm eq}/T_{\rm b}$ versus $T_{\rm eq}$. Note that $T_{\rm b}$ is systematically larger than $T_{\rm eq}$, which implies that the measured planet brightness temperature cannot be explained only by its equilibrium temperature from the stellar radiation field, assuming the blackbody model is correct.  This suggests that reflection may not be negligible for many of these planets.  This is not surprising, considering that many hot Jupiters are known to have clouds or hazes \citep[e.g.,][]{sing2016}.  Recent work by, e.g., \citet{schwartz2015}, has also demonstrated that reflected light can have a significant impact, even in the near-infrared.  Therefore, our findings are consistent with other recent studies of exoplanet atmospheres. Moreover, this difference appears to be larger at lower temperatures, where the thermal flux become smaller, and therefore is less dominant over other sources. This corroborates the hypothesis that reflected light may have a significant impact in the measured near-infrared brightness, especially at lower equilibrium temperatures. 

\begin{table*}
\caption{Parameters of exoplanet systems with existing K-band detections. The parameters of each system were obtained from the Open Exoplanet Catalogue (http://www.openexoplanetcatalogue.com) through the astroquery package using the library open\_exoplanet\_catalogue. The planet-to-star flux ratios for the first eight objects were obtained from our measurements and the remaining objects were obtained from \citet{zhou2015}.}
\label{table:exosystems}
\begin{tabular}{lcccccccccc}
\hline
ID & flux ratio &  $M_{*}$ &  $R_{*}$  &  $T_{*}$ &  K  &  [Fe/H] &  $M_{p}$ &  $R_{p}$& $a_{p}$ & age \\
& $f_{p}/f_{*}$ [\%]  &  [$M_{\odot}$] &  [$R_{\odot}$]  &  [K] &  [mag]  &  [dex] &  [$M_{J}$] &  [$R_{J}$]& [AU] & [Gyr] \\
\hline
\hline
WASP-12  & $0.323\pm0.011$ & 1.35 & 1.599 & 6118.0 & 10.188 & 0.3 & 1.404 & 1.736 &  0.02293 & 1.7\\
KELT-4A  & $0.220\pm0.040$ & 1.128 & 1.495 & 6140.0 & 8.689 & -0.163 & 0.83 & 1.586 &  0.04228 & 4.38\\
KELT-7   & $0.040\pm0.012$ & 1.535 & 1.732 & 6789.0 & 7.543 & 0.139 & 1.28 & 1.533 & 0.04415 & 1.3\\
KELT-2A  & $0.118\pm0.017$ & 1.314 & 1.836 & 6148.0 & 7.346 & -0.034 & 1.524 & 1.29 &  0.05504 & -\\
HAT-P-33 & $0.153\pm0.022$ & 1.403 & 1.777 & 6401.0 & 10.004 & 0.05 & 0.763 & 1.827 &  0.0503 & 2.4\\
Kepler-5 & $0.080\pm0.025$ & 1.374 & 1.793 & 6297.0 & 11.769 & 0.04 & 2.114 & 1.431 &  0.05064 & -\\
WASP-31  & $0.102\pm0.017$ & 1.16 & 1.24 & 6200.0 & 10.65 & -0.19 & 0.478 & 1.537 &  0.04657 & -\\
TrES-4   & $0.202\pm0.090$ & 1.388 & 1.798 & 6200.0 & 10.33 & 0.14 & 0.494 & 1.838 & 0.0516 & 2.9 \\
WASP-14  & $0.172\pm0.025$ & 1.211 &  1.306 & 6462.0 & 8.621 & 0.0 & 7.34 & 1.281 & 0.0360 & 0.75 \\
\hline
WASP-48 & 0.11 $\pm$ 0.03 & 1.19 & 1.75 & 5920.0 & 10.372 & -0.12 & 0.98 & 1.67 & 0.03444 & -\\
OGLE-TR-113 & 0.17 $\pm$ 0.05 & 0.779 & 0.774 & 4790.0 & 13.0 & 0.09 & 1.26 & 1.051& 0.02288 & 0.7\\
WASP-43 & 0.19 $\pm$ 0.03 & 0.717 & 0.667 & 4520.0 & 9.267 & -0.01 & 2.034 & 1.036 &  0.01526 & 0.4\\
WASP-46 & 0.25 $\pm$ 0.06 & 0.956 & 0.917 & 5620.0 & 11.401 & -0.37 & 2.101 & 1.31 &  0.02448 & 1.4\\
CoRoT-1 & 0.28 $\pm$ 0.07 & 0.95 & 1.11 & 6298.0 & 12.149 & -0.3 & 1.03 & 1.49 &  0.0254 & -\\
CoRoT-2 & 0.16 $\pm$ 0.09 & 0.97 & 0.902 & 5575.0 & 10.31 & 0.0 & 3.31 & 1.465 &  0.0281 & -\\
WASP-18 & 0.15 $\pm$ 0.02 & 1.24 & 1.23 & 6400.0 & 8.131 & 0.16 & 10.43 & 1.165 &  0.02047 & 0.63\\
WASP-19 & 0.37 $\pm$ 0.07 & 0.97 & 0.99 & 5500.0 & 10.481 & 0.02 & 1.168 & 1.386 &  0.01655 & 11.5\\
WASP-36 & 0.14 $\pm$ 0.04 & 1.02 & 0.943 & 5881.0 & 11.294 & -0.31 & 2.279 & 1.269 & 0.02624 & 3.0\\
WASP-10 & 0.14 $\pm$ 0.02 & 0.71 & 0.783 & 4675.0 & 9.983 & 0.03 & 3.06 & 1.08 &  0.0371 & 0.8\\
WASP-76 & 0.14 $\pm$ 0.04 & 1.46 & 1.73 & 6250.0 & 8.243 & 0.23 & 0.92 & 1.83 &  0.033 & -\\
TrES-3 & 0.24 $\pm$ 0.04 & 0.88 & 0.85 & 5720.0 & 10.608 & 0.001 & 1.91 & 1.305 & 0.0226 & -\\
TrES-2 & 0.06 $\pm$ 0.01 & 0.98 & 1.0 & 5850.0 & 9.846 & -0.15 & 1.253 & 1.169 & 0.03556 & 5.1\\
HAT-P-23 & 0.23 $\pm$ 0.05 & 1.13 & 1.203 & 5905.0 & 10.791 & 0.15 & 2.09 & 1.368 &  0.0232 & 4.0\\
KELT-1 & 0.16 $\pm$ 0.02 & 1.335 & 1.471 & 6516.0 & 9.437 & 0.052 & 27.38 & 1.116 &  0.02472 & 1.75\\
Qatar-1 & 0.14 $\pm$ 0.03 & 0.85 & 0.8 & 4910.0 & 10.409 & 0.2 & 1.33 & 1.164 & 0.02343 & 4.0\\
Kepler-13 & 0.12 $\pm$ 0.05 & 2.05 & 2.55 & 8500.0 & 9.425 & -0.14 & 9.28 & 1.51 &  0.03423 & -\\
HAT-P-32 & 0.18 $\pm$ 0.06 & 1.176 & 1.387 & 6001.0 & 9.99 & -0.16 & 0.941 & 2.037 &  0.0344 & 3.8\\
WASP-33 & 0.27 $\pm$ 0.04 & 1.495 & 1.444 & 7400.0 & 7.468 & 0.1 & 4.59 & 1.438 &  0.02558 & -\\
WASP-3 & 0.18 $\pm$ 0.02 & 1.24 & 1.31 & 6400.0 & 9.361 & 0.0 & 2.06 & 1.454 &  0.0313 & -\\
WASP-4 & 0.19 $\pm$ 0.01 & 0.93 & 1.15 & 5500.0 & 10.746 & -0.03 & 1.1215 & 1.363 &  0.02312 & -\\
WASP-5 & 0.27 $\pm$ 0.06 & 1.0 & 1.084 & 5700.0 & 10.598 & 0.09 & 1.637 & 1.171 &  0.02729 & 3.0\\
HAT-P-1 & 0.11 $\pm$ 0.03 & 1.133 & 1.115 & 5975.0 & 8.858 & 0.13 & 0.524 & 1.217 &  0.05535 & 3.6\\
\hline
\end{tabular}
\end{table*}

\section{Conclusions}
\label{sec:conclusions}

We have presented high precision photometric time series in the near infrared for eight transiting exoplanets during their predicted secondary eclipses. A thorough investigation of the experimental design and reduction steps is presented. We have performed photometric measurements using optimized aperture sizes, which are calculated automatically based on the flux fraction and signal-to-noise ratio collected within the aperture. We obtained reliable differential photometry with precision as low as $0.11\%$. We have also presented a robust approach for combining the differential light curves without being biased by the presence of the eclipse. This allowed us to remove systematics and to obtain a final light curve with improved signal-to-noise. We demonstrated our data reduction technique by analyzing the secondary eclipse of WASP-12b, which has been previously analyzed and published by \citet{croll2015}. For some of our targets, the uncertainties on the measured eclipse depths are likely to be underestimated due to residual red noise which has not been accounted for in our analysis.
We have presented a Bayesian analysis of our data, where we applied the \citet{goodmanweare2010} affine invariant MCMC ensemble sampler to measure the eclipse depths. Our measurements have been analyzed along with other results from previous measurements of eclipses in the K-band.  The measurements presented here increased the sample of exoplanets with published eclipses in K-band by 35\%, thereby producing the most complete sample to date of exoplanets with detected planet-to-star flux ratios in the same near infrared band pass.
We investigated the full sample of exoplanets with measured eclipses in the K-band. We compared the detected eclipse depths to the expected depths from a simple model for a planet emitting as a blackbody in thermal equilibrium with the stellar radiation field, where we considered both a complete heat redistribution model and a no-heat redistribution model. The brightness temperatures obtained from the eclipse depths present an excess compared to the equilibrium temperatures. The excess in the brightness temperatures appear to be larger at lower equilibrium temperatures, which suggests that another source of radiation (e.g., reflected light and/or internal heat) has a significant contribution to the near-infrared flux measured from hot Jupiters. 

\section*{Acknowledgements}

Based on observations obtained with WIRCam, a joint project of CFHT, Taiwan, Korea, Canada, France, and the Canada-France-Hawaii Telescope (CFHT) which is operated by the National Research Council (NRC) of Canada, the Institut National des Sciences de l'Univers of the Centre National de la Recherche Scientifique of France, and the University of Hawaii. 

This research has made use of the Brazilian time at CFHT through the agreement between the Brazilian Ministry of Science Technology Innovation and Communications (MCTIC) and the CFHT. E. Martioli acknowledges the support of CNPq, under the Universal grant process number 443557/2014-4, and also the support of FAPEMIG, under project number 01/2014-23092. This research has also made use of: the SIMBAD database, which is operated at CDS, Strasbourg, France; of Astropy, a community-developed core Python package for Astronomy \citep{astropy2013}. The authors appreciate the work done by the CFHT staff and in particular to the QSO team for scheduling the observations, for promptly responding to our requests, and for ensuring that these observations were accomplished successfully. The authors also wish to recognize and acknowledge the very significant cultural role and reverence that the summit of Mauna Kea has always had within the indigenous Hawaiian community.  We are most fortunate to have had the opportunity to conduct observations from this mountain. D.A. acknowledges support by an appointment to the NASA Postdoctoral Program at Goddard Space Flight Center, administered by USRA through a contract with NASA. P.A.W acknowledges the support of the French Agence Nationale de la Recherche (ANR), under program ANR-12-BS05-0012 "Exo-Atmos". Work by T.G.B., B.S.G. and D.J.S. was partially supported by NSF CAREER Grant AST-1056524.







\newpage 

\bsp	
\label{lastpage}
\end{document}